\documentclass[letterpaper,titlepage,11pt]{article}
\pdfoutput=1
\usepackage{jheppub}
\usepackage{hyperref}
\usepackage{epsfig}
\usepackage{graphicx}
\usepackage{amsmath}
\usepackage{amsfonts}
\usepackage{amssymb}
\usepackage{mathrsfs}
\usepackage{fancybox}
\usepackage{ragged2e}
\usepackage{tikz}
\usepackage{wasysym}

\newcommand{\mycite}[1]{~{\cite{#1}}}
\newcommand{\myfigref}[1]{~{Fig.~(\ref{#1})}}
\newcommand{\myref}[1]{~{(\ref{#1})}}
\newcommand{\tr}{\mathop{{\rm Tr}}}
\newcommand{\beq}{\begin{equation}}
\newcommand{\eeq}{\end{equation}}

\newcommand{\be}{\begin{equation*}}
\newcommand{\ee}{\end{equation*}}
\title{Three-point functions of BMN operators at weak and strong
coupling II. One loop matching.}
\author[a]{Gianluca Grignani,}
\author[a,b]{A.~V.~Zayakin}
\affiliation[a]{Dipartimento di Fisica, Universit\`a di Perugia,\\
I.N.F.N. Sezione di Perugia,\\
Via Pascoli, I-06123 Perugia, Italy}
\affiliation[b]{Institute of Theoretical and Experimental Physics,\\
B.~Cheremushkinskaya ul. 25, 117259 Moscow, Russia}
\emailAdd{grignani@pg.infn.it}
\emailAdd{a.zayakin@gmail.com}
\abstract{In
a previous paper [JHEP 06 (2012) 142] we have shown that the
fully dynamical three-point correlation functions of BMN operators
are identical at the tree level in the planar limit of
perturbative field theory and, on the string theory side,
calculated by means of the Dobashi-Yoneya three string vertex in
the Penrose limit. Here we present a one-loop calculation of the
same quantity both on the field-theory and string-theory side,
where a complete identity between the two results is demonstrated.}

\begin{document}
\maketitle

\setcounter{page}{1}

\section{Introduction}
Three-point correlators in gauge/gravity duality have become a
very hot topic in the last couple of years. A variety of
fundamental results for the three-point functions have been
obtained by means of string theory quasiclassics in the strong
coupling
limit\mycite{Buchbinder:2010vw,Costa:2010rz,Roiban:2010fe,
Zarembo:2010rr,Buchbinder:2011jr,Michalcik:2011hh,Klose:2011rm,
Georgiou:2010an,Hernandez:2010tg,
Arnaudov:2010kk,Park:2010vs,Russo:2010bt,Bak:2011yy,
Arnaudov:2011wq,Hernandez:2011up,Bai:2011su,
Ahn:2011zg,Arnaudov:2011ek,Janik:2011bd}, exact string theory
Hamiltonian in the Penrose limit, Bethe Ansatz at strong and weak
coupling\mycite{Escobedo:2011xw,Escobedo:2010xs,
Gromov:2011jh,Gromov:2012vu,Serban:2012dr,Kostov:2012jr,
Grignani:2012yu}, perturbative large-N field
theory\mycite{Okuyama:2004bd,Alday:2005nd}, and comparison between
weakly coupled planar field theory and results in semiclassical
string
theory\mycite{Georgiou:2012zj,Bissi:2011ha,Bissi:2011dc,Foda:2011rr}.

Comparing the gravity side and the gauge side directly is possible
only under very specific asymptotic conditions such as,
\textit{e.g.}, the Frolov--Tseytlin limit\mycite{Frolov:2003xy}.
In this limit it is the smallness of
$\lambda^\prime\equiv\frac{\lambda}{J^2}$, where $\lambda$ is the
't Hooft coupling, and $J$ is an R-charge,  that permits a
comparison. The limit is, in fact, compatible with both small and
large $\lambda$. Thus, if for example one deals with a small
number of impurities, one can calculate the correlator on the
field theory side where $\lambda\to 0$, and at the same time use
the full string interaction Hamiltonian in the Penrose limit,
where the coupling is large, $\lambda\to \infty$, to obtain
legitimately comparable results. In this paper we will be looking
for the $\lambda^\prime$ corrections to three-point correlation
functions for states with few excitations, both at weak and at
strong coupling.

It has been discussed at length in\mycite{Bissi:2011ha} whether
the one-loop corrections for three-point functions are expected to
match, even at the leading order. The limits  $\lambda\gg 1 $ and
$\lambda\ll 1 $ are normally incompatible, so that even the
tree-level matching is not necessarily to be
expected~\cite{Escobedo:2010xs,Georgiou:2010an}. In the case of
2-point functions, it is well-known that there is an agreement
between the anomalous dimension of certain gauge theory operators
and the energy of the dual string states directly computed from
the string sigma model. In\mycite{Harmark:2008gm}, an argument for
the seemingly coincidental matching up to and including the
one-loop correction was provided for states in the $SU(2)$ sector
of $\mathcal{N}=4$ SYM.    The argument of\mycite{Harmark:2008gm}
goes as follows: consider a near-BPS state, with $E-J\ll 1$ and
$J\gg 1$. This limit is feasible both on the string side and on
the field theory side. On the string theory side quantum
corrections to the semiclassical configuration become suppressed
and the limit remains valid. However, it was also discussed
in\mycite{Bissi:2011ha} that this argument, being perfectly valid
for the two-point functions, does not apply to the three-point
functions, thus there will
generally be gauge-field terms of order $\frac{\lambda^0}{J^2}$,
whereas string theory yields $\frac{\lambda}{J^2}$. Thus one-loop
matching is in general not expected and in fact it is not found in
a particular example of a heavy-heavy light three-point
function\mycite{Bissi:2011ha}. As far as we know, this has been
the only example of an explicit one-loop comparison between
perturbative field theory and semiclassical string theory so far.
The mismatch might however be due to the difficult identification
of the two loop corrected gauge theory states.

In\mycite{Gromov:2012vu} a general formula for three point
correlators of single trace operators  with arbitrary number of
impurities $N_i$, that satisfy $N_1=N_2+N_3$,  is provided at one-loop. Two operators are long (and highly
excited) and the other is shorter. An amazing matching
is seen numerically in the limit of $N_3\to \infty$, where the
semiclassical calculation fully conforms to the Bethe Ansatz
calculation.

In this paper we perform an explicit one-loop check of the
matching in a different sector, where the operator-state
identification between gauge and string theory is perfectly
well-defined~\cite{Berenstein:2002jq}. The main object of our
analysis are two-magnon BMN operators
\beq\mathcal{O}_{ij,n}^{J}=\frac{1}{\sqrt{J
N^{J+2}}}\sum^J_{l=0}\tr \left(\phi_i Z^l\phi_j
Z^{J-l}\right)\psi_{n,l},\eeq which fall into the three
irreducible representations of $SO(4)$
\beq \bf 4\otimes 4=1\oplus 6 \oplus 9,\eeq
where $\mathbf 1$ is the trace (T), $\mathbf 6 $ is the
antisymmetric (A), $\mathbf 9$ is the symmetric traceless
representation (S). The wave-functions for different
representations are
\beq\begin{array}{l} \psi^S_{n,l}=\cos\frac{(2l+1)\pi n}{J+1},\\
\\
\psi^A_{n,l}=\sin\frac{2(l+1)\pi n}{J+2},\\ \\
\psi^T_{n,l}=\cos\frac{(2l+3)\pi n}{J+3}.\end{array} \eeq

We consider three operators:
$\mathcal{O}_1=\mathcal{O}^{J_1,12}_{n_1},
\mathcal{O}_2=\mathcal{O}^{J_2,23}_{n_2},
\mathcal{O}=\mathcal{O}^{J,31}_{n}$, where $n_1,n_2,n_3$ are the
operator momenta, $J_1,J_2,J_3$ are their R-charges $R_3$,
$J=J_1+J_2$, $J_1=J y$,$J_2=J(1-y)$.
The flavor indices are chosen as $(12),(23),(31)$ to represent the
symmetric sector of the theory. The symmetric states are more
interesting for our analysis since they provide a non-trivial test
of the calculation by requiring cancellation of the $J^2$ and $J$
order of the one-loop correction. Unlike the trace state, the
traceless symmetric state is advantageous since one avoids the
complications with subtraction of $\tr Z^{J+1} \bar{Z}$. Our
operators are orthonormalized at tree level, therefore the
correlator coincides with the structure constant.
We shall be looking for the quantity
\beq C_{123}=\langle \bar{\mathcal{O}}_3
\mathcal{O}_1\mathcal{O}_2\rangle\eeq as a function of
$y,J,n_1,n_2,n_3$. This gives the three point correlator of  these
operators, thanks to the conformal invariance of $\mathcal{N}=4$
SYM.
 We shall calculate this correlator on the field
theory side as well as on the string theory side. In our previous
work\mycite{Grignani:2012yu} we have already shown that at tree
level these correlators do coincide with the corresponding
quantities computed from the string side. Now we proceed to derive
the one-loop contributions.

\section{String theory calculation}
In terms of the BMN basis $\{\alpha_m\}$ the operators in question
look like states
\beq \mathcal{O}_m=\alpha_m^\dagger\alpha_{-m}^\dagger |0\rangle
\eeq
The three-point function is related to the matrix element of the
Hamiltonian as follows
\beq \langle\bar{\mathcal{O}}_3\mathcal{O}_1\mathcal{O}_2\rangle=
\frac{4\pi}{-\Delta_3+\Delta_1+\Delta_2}\sqrt{\frac{J_1
J_2}{J}}H_{123}\eeq where
\beq\begin{array}{l} \displaystyle
\Delta_{1}=J_1+2\sqrt{1+\lambda' n_1^2},\\
\\ \displaystyle \Delta_{2}=J_2+2\sqrt{1+\lambda' n_1^2},\\ \\ \displaystyle
\Delta_{3}=J+2\sqrt{1+\lambda' n_3^2},
\end{array}
\eeq
and the matrix element is defined as
\beq H_{123}=\langle 123 |V\rangle.\eeq
There has  been some ambiguity in the literature with regard to
how the proper prefactor in the vertex function  $V$ in the
pp-wave looks like\mycite{Spradlin:2002rv,
Dobashi:2002ar,Pearson:2002zs,Pankiewicz:2002tg,
He:2002zu,Pankiewicz:2003ap,DiVecchia:2003yp,
Dobashi:2004nm,Dobashi:2004ka,Shimada:2004sw,Lee:2004cq,
Grignani:2005yv,Grignani:2006en}; we use the findings
of\mycite{Grignani:2006en} to start with the Dobashi--Yoneya
prefactor~\cite{Dobashi:2004nm} in the natural string basis
$\{a_m^r\}$.
\beq V=Pe^{\frac{1}{2}\sum_{m,n}
N^{rs}_{mn}\delta^{IJ}a^{rI\dagger}_m a^{sJ\dagger}_n}. \eeq
Here $I,J$ are $SU(4)$ flavour indices, $r,s$ run within $1,2,3$
and refer to the first, second and third operator. The natural
string basis is related to the BMN basis for $m>0$ as follows
\beq \left\{\begin{array}{l}\displaystyle
\alpha_m=\frac{a_m+ia_{-m}}{\sqrt{2}}\\ \\ \displaystyle
\alpha_{-m}=\frac{a_m-ia_{-m}}{\sqrt{2}}
\end{array},\right.
\eeq
\noindent The Neumann matrices are given as\mycite{He:2002zu}
\beq\begin{array}{l}\displaystyle
 N_{m,n}^{rs}=\frac{1}{2\pi}
\frac{(-1)^{r(m+1)+s(n+1)}}{x_s\omega_{rm}+x_r\omega_{sn}}
\sqrt{\frac{x_rx_s(\omega_{rm}+\mu x_r)(\omega_{sn}+\mu
x_s)s_{rm}s_{qn}}{\omega_{rm}\omega_{sn}}},\\ \\ \displaystyle

N_{-m,-n}^{rs}=-\frac{1}{2\pi}
\frac{(-1)^{r(m+1)+s(n+1)}}{x_s\omega_{rm}+x_r\omega_{sn}}
\sqrt{\frac{x_rx_s(\omega_{rm}-\mu x_r)(\omega_{sn}-\mu
x_s)s_{rm}s_{qn}}{\omega_{rm}\omega_{sn}}}, \end{array}\eeq where
$m,n$ are always meant positive,
\beq\begin{array}{l} s_{1m}=1,\\
s_{2m}=1,\\
s_{3m}=-2\sin(\pi m y),\\
\end{array}\eeq
and
\beq\begin{array}{l} x_{1}=y,\\
x_{2}=1-y,\\
x_{3}=-1,\\
\end{array}\eeq
the frequencies of the string oscillators are then given as
\beq \omega_{r,m}=\sqrt{m^2+\mu^2 x_r^2},\eeq
and the parameter $\mu$ is directly related to the Frolov-Tseytlin
expansion parameter $\lambda^\prime$
\beq \mu=\frac{1}{\sqrt{\lambda^\prime}}.\eeq
The one-loop calculation of the correlation function will amount a
next-order expansion in $\frac{1}{\mu^2}$  of the matrix element.
An essential feature of the Dobashi-Yoneya prefactor we are using
is that the prefactor is supported with positive modes only
\beq P=\sum_{m>0}\sum_{r,I} \frac{\omega_r}{\mu \alpha_r}a^{I
r\dagger}_m a^{I r}_m.\eeq
Due to the flavour structure of $C_{123}$ the only combinations of
terms from the exponent that could contribute are $N^{12}_{n_1n_2}
N^{23}_{n_2n_3} N^{31}_{n_3n_1}$. The leading order contribution
is
\beq\displaystyle  C^0_{123}=\frac{1}{\pi^2} \frac{\sqrt{J}}{N}
\frac{n_3^2 y^{3/2}(1-y)^{3/2}\sin^2(\pi n_3 y) }{(n_3^2y^2
-n_1^2)(n_3^2(1-y)^2 -n_2^2)}\eeq
The overall factor $-4$ difference with \mycite{Grignani:2012yu}
is due to wave-function normalization. The next-order coefficient
in the expansion
\beq C_{123}=C^0_{123}\left(1+\lambda^\prime
c^{1}_{123}\right),\eeq where $c^{1}_{123}\equiv
\frac{C^1_{123}}{C^0_{123}}$ is
\beq c^1_{123}=-\frac{1}{4}\left(\frac{n_1^2}{y^2}
+\frac{n_2^2}{(1-y)^2}+n_3^2 \right).\eeq
 Let us compare this
calculation to the field theory calculation.

\section{Planar Field Theory at One Loop}
\begin{figure}[h!]
\begin{center}
\includegraphics[height =5cm, width=9cm]{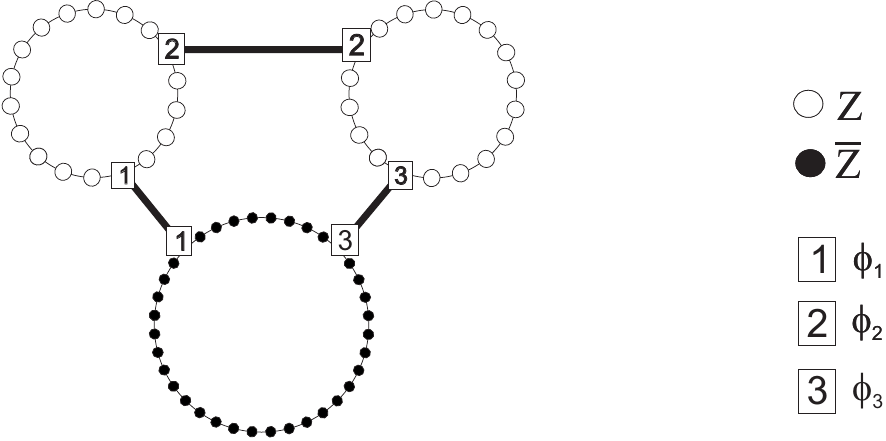}
\caption{\label{leadingorder}Leading order  diagram for the
three-point correlation function of the fully dynamic BMN
operators from the symmetric traceless sector.}
\end{center}
\end{figure}

To calculate the correlation function at one loop level we use the
technique developed in \mycite{Kristjansen:2002bb,Beisert:2002bb}.
The leading order correlation function was calculated by us in our
preceding paper~\cite{Grignani:2012yu} and is given by the
diagram of\myfigref{leadingorder}.

\noindent Only $\phi$ propagators are shown explicitly in the
figure; planarity is imposed, the  $Z$ fields do not have any
extra choice than to contract a definite $\bar{Z}$, thus the
diagram contributes exactly once. The diagram\myref{leadingorder}
refers to the correlation function calculated from planar
perturbation theory and is related to the $C_{123}$ defined above
\beq S= N\sqrt{J_1 J_2 J} C_{123} \eeq
which is given in general by a $\lambda^\prime$ expansion
\beq S=S^{(0)}+S^{(1)}+O(\lambda^{\prime 2})\eeq
and evaluates in the leading order to \beq
S^{(0)}=\sum_{l_1=0,l_2=0}^{l_1=J_1,L_2=J_2} \cos \frac{\pi(2
l_1+1)}{J_1+1} \cos \frac{\pi(2 l_2+1)}{J_2+1} \cos \frac{\pi(2
(l_1+l_2)+1)}{J+1},\eeq
which after the $1/J$ expansion and the due normalization of the
operator to unity yields
\beq C^0_{123}=\frac{1}{\pi^2}\frac{\sqrt{J}}{N} \frac{n_3^2
y^{3/2}(1-y)^{3/2}\sin^2(\pi n_3 y) }{(n_3^2y^2
-n_1^2)(n_3^2(1-y)^2 -n_2^2)},\eeq corresponding exactly to the
result above.

\begin{figure}
\begin{center}
\includegraphics[height=18.5cm, width=16cm]{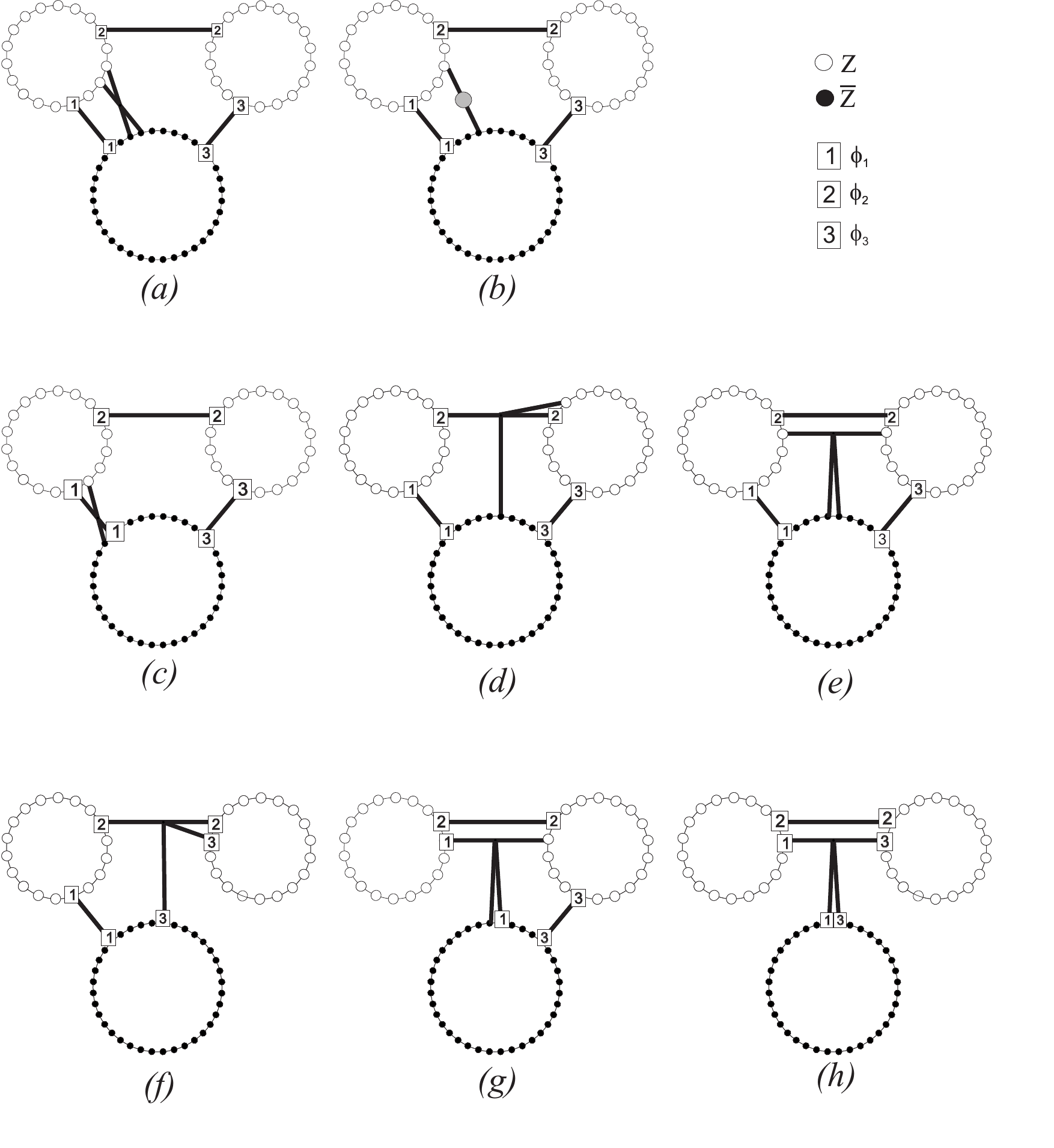}
\caption{\label{eight}Eight diagrams that contribute to the
three-point correlation function of the fully dynamic BMN
operators from the symmetric traceless sector $C_{123}$ at
one-loop level.}
\end{center}
\end{figure}

At the one loop level we estimate the next-order terms in the
$\lambda^\prime$ expansion $S^{(1)}$ considering all possible
insertions of the interaction terms
\beq H_2=\frac{\lambda}{8\pi^2}\left(I - P\right) \eeq
into\myfigref{leadingorder}, where $I$ is unity operator and $P$
permutation operator, both acting on nearest neighbors.  It is
convenient to split the Hamiltonian into the part $P$ and the part
$I$; in the diagrams below the four-point vertices are meant as
pure permutations, i.e. acting as $P$ solely. The set of the
resulting eight diagrams that contribute to $C_{123}^1$ are shown
in\myfigref{eight}. Only the insertions and the $\phi^{1,2,3}$
propagators are shown; the configuration of the rest of the
propagators is fully determined by planarity rules.

Three types of mixing aggravate our task: the admixture of
multi-trace operators (eq. (3.14) in\mycite{Beisert:2002bb}), magnon
mode number non-conserving admixture caused by the coupling dependent wave function correction
(BMN operator redefinition, eq.
(5.18) in\mycite{Beisert:2003tq}), and the admixture with fermionic
operators\mycite{Georgiou:2008vk,Georgiou:2009tp},

Multi-trace operator redefinition is organized as
\beq \mathcal{O}^{J,12
\prime}_{n}=\mathcal{O}^{J,12}_{12,n}-\frac{J^2}{N}\sum_{k,r}
\frac{r^{3/2}\sqrt{1-r}\sin^2(\pi n r)k}{\sqrt{J}\pi^2
(k-nr)^2(k+nr)}\mathcal{T}_{12,k}^{J,r} \eeq
where $\mathcal{T}_{12,n}^{J,r}=\mathcal{O}_{n}^{rJ,12}
\mathcal{O}^{(1-r)J}$, $\mathcal{O}^J$ being the normalized vacuum
operator of length $J$. For our kinematics the multi-trace mixing
becomes significant only in the next-order corrections in $1/N$.

The magnon mode number nonpreserving BMN operator $\mathcal{O}_n$
redefinition in the order $\lambda^\prime$ is organized as
\beq \mathcal{O}^{J,12  \prime}_{n}=\mathcal{O}^{J,12 }_{n}-
\frac{\lambda}{(J+1)\pi^2}\sum^{[J/2]}_{m=1}\delta_{m\neq n}
\frac{\sin^2\frac{\pi n}{J+1}\cos\frac{\pi n}{J+1}\sin^2 \frac{\pi
m}{J+1}\cos\frac{\pi m}{J+1}}{\sin^2\frac{\pi
n}{J+1}-\sin^2\frac{\pi m}{J+1}} \mathcal{O}_{m}^{J,12} ,\eeq
here $[J/2]$ denotes the integer part of $J/2$. This operator
redefinition has been considered by us and has been shown not to
contribute due to suppression by higher-order powers of
$\frac{1}{J}$.

The admixture with fermionic operators is the most difficult to
handle. At order $\lambda$ it is not yet known for the class
of symmetric traceless operators considered in this work. The
mixing for the trace class operators is derived in eq. (2.1)
in\mycite{Georgiou:2009tp}. If the mixing for the symmetric traceless
sector was described by a formula of the same type as eq. (2.1) in
\mycite{Georgiou:2009tp} (which is still to be determined whether
it is so or not), a rough estimate yields that the mixing might contribute in our case at the order $g^2/J^2$. However,
the tree-level contribution is of order $J^2$ and the one-loop goes as $g^2\,J^0$. Thus the mixing correction will appear at the next $1/J^2$
order while holding $\lambda^\prime$ order fixed, so that it would not contribute.

Anyway, since we find complete identity between the string and gauge theory
calculations this is a clear sign that at the given orders in
$\lambda^\prime$ and $\frac{1}{J}$ no extra mixing has to be taken
into account. Whatever the
mixing is, presumably it takes place both on the string and field theory sides and gives the same contribution, so that the
coincidence of the results might be accounted for.
Surely this issue deserves further investigation.

Some comments on the classification and evaluation of these
diagrams are necessary. They all arise from  expressions with
three sums $\sum_{l_1=0,l_2=0,l_3=0}^{J_1,J_2,J}$ over the three
wave functions $\psi_{n_1 l_1},\psi_{n_2 l_2},\psi_{n_3 l_3}$. The
diagrams are done in the planar limits, thus only combinations
with non-intersecting propagators are being considered. One of the
sums is always lifted by conservation law. The diagrams $(a)$ and
$(b)$ scale as $J^3$. The diagrams $(c), (d), (e)$ contribute as
$J^2$, $(f), (g)$ scale as $J$ and $(h)$ scales as $J^0$.

In cases $(a)$ and $(b)$ the self energy of $Z$ or $\phi$, or the
$Z^4$ scattering insertion produce an extra factor: $l_1$ and $l_2$
fixed, the insertion can be inserted into $\sim J$ locations
without breaking the planarity. This raises the order of the
diagrams up to $J^3$; happily their leading order  terms do cancel
between themselves, and the remaining $J^2$ and $J$ terms cancel
with the rest. The other diagrams do not give rise to extra factors. Also
note that while $(c), (d), (e)$ can be realized with arbitrary $l_1,l_2$,
the $(f), (g)$ exist only for marginal cases, with either $l_1=0$ or
$l_2=0$ (we do not show in\myfigref{eight} the diagrams that
differ from $(a)-(h)$ by the $1\leftrightarrow2$ symmetry only).
We systematize these contributions in the table\myref{tab1} below.
Notice that in principle we could have included explicitly the
scheme dependence into the ``two-point''-type diagrams $(a)-(c)$
(where the contribution of the operator containing impurities of
type 2 and 3 can be factor out) and the ``three-point''-type
diagrams $(d)-(h)$. The two groups of diagrams are in fact scheme
dependent, however the scheme dependence cancels exactly once the
two groups are added together. Such cancellation, that leads to
the scheme-independence of the full result, is the same as the one
discussed in\mycite{Okuyama:2004bd} and \mycite{Alday:2005nd}.
\begin{table}[h!]\caption{Classification of diagrams at
one-loop level for the three-point correlator of BMN operators
\label{tab1}}
\begin{equation*}
\begin{array}{|l|l|l|l|l|}\hline\hline
\mbox{Diagram}&\mbox{Vertex
type}&\mbox{Order}&\mbox{Coefficient}\\ \hline\hline a&Z^4
&J^3&1\\ \hline b&\mbox{Self-energy}&J^3&-1\\ \hline
c&Z^2\phi^2&J^2&1\\ \hline d&Z^2\phi^2&J^2&1/2\\ \hline
e&Z^4&J^2&1/2\\ \hline f&\phi^4&J&1\\ \hline g&Z^2\phi^2&J&1\\
\hline h&\phi^4&1&1\\ \hline\hline
\end{array}
\end{equation*}
\end{table}

These diagrams are evaluated as
\be
\begin{array}{l}\displaystyle
S^{(1)}_{a}=\sum_{l_1=2}^{l_1=J_1}\,\sum_{l_2=0}^{l_2=J_2}
(l_1-1)\cos \frac{\pi n_1(2 l_1+1)}{J_1+1}\cos
\frac{\pi n_2(2 l_2+1)}{J_2+1} \cos \frac{\pi n_3(2 (l_1+l_2)+1)}{J+1}+\\
\\ \displaystyle \hphantom{S^1_{a}=}
 +\sum_{l_1=0}^{l_1=J_1-2}\,\,\sum_{l_2=0}^{l_2=J_2} (J_1-l_1-1)\cos
\frac{\pi n_1(2 l_1+1)}{J_1+1}\cos \frac{\pi n_2 (2 l_2+1)}{J_2+1}
\cos \frac{\pi n_3(2 (l_1+l_2)+1)}{J+1}
+(1\leftrightarrow 2),\\ \\
\displaystyle
S^{(1)}_{b}=(J+3)\sum_{l_1=0}^{l_1=J_1}\sum_{l_2=0}^{l_2=J_2} \cos
\frac{\pi n_1(2 l_1+1)}{J_1+1}\cos \frac{\pi n_2(2 l_2+1)}{J_2+1}
\cos \frac{\pi n_3(2 (l_1+l_2)+1)}{J+1},\\ \\ \displaystyle
\end{array}\ee
\be
\begin{array}{l}\displaystyle
S^{(1)}_{c}=\sum_{l_1=0}^{l_1=J_1-1} \sum_{l_2=0}^{l_1=J_2} \cos
\frac{\pi n_1(2 l_1+1)}{J_1+1} \cos \frac{\pi n_2(2 l_2+1)}{J_2+1}
\cos \frac{\pi n_3 (2
(l_1+l_2+1)+1)}{J+1}+\\ \\
\displaystyle \hphantom{S^1_{a}=} +\sum_{l_1=0}^{l_1=J_1}
\sum_{l_2=0}^{l_2=J_2-1} \cos \frac{\pi n_1 (2 l_1+1)}{J_1+1} \cos
\frac{\pi n_2(2 l_2+1)}{J_2+1} \cos \frac{\pi n_3(2
(l_1+l_2+1)+1)}{J+1}+
\\ \\ \displaystyle \hphantom{S^1_{a}=}
+\sum_{l_1=1}^{l_1=J_1} \sum_{l_2=0}^{l_2=J_2} \cos \frac{\pi
n_1(2 l_1+1)}{J_1+1} \cos \frac{\pi n_2(2
l_2+1)}{J_2+1} \cos \frac{\pi n_3(2 (l_1+l_2-1)+1)}{J+1} \\ \\
\displaystyle \hphantom{S^1_{a}=} +\sum_{l_1=0}^{l_1=J_1}
\sum_{l_2=1}^{l_1=J_2} \cos \frac{\pi n_1(2 l_1+1)}{J_1+1} \cos
\frac{\pi n_2(2
l_2+1)}{J_2+1} \cos \frac{\pi n_3 (2 (l_1+l_2-1)+1)}{J+1},\\ \\
\displaystyle S_d^{(1)}=S_c,\\ \\ \displaystyle  \end{array}\ee
\be
\begin{array}{l}\displaystyle
S_e^{(1)}=\sum_{l_1=1}^{l_1=J_1}\sum_{l_2=1}^{l_2=J_2} \cos
\frac{\pi n_1(2 l_1+1)}{J_1+1} \cos \frac{\pi n_2 (2
l_2+1)}{J_2+1} \cos \frac{\pi n_3 (2 (l_1+l_2)+1)}{J+1}+ \\ \\
\displaystyle \hphantom{S^1_{a}=} +
\sum_{l_1=0}^{l_1=J_1-1}\sum_{l_2=0}^{l_2=J_2-1} \cos \frac{\pi
n_1 (2 l_1+1)}{J_1+1}\cos \frac{\pi n_2 (2 l_2+1)}{J_2+1} \cos
\frac{\pi n_3 (2
(l_1+l_2)+1)}{J+1},\\ \\
\displaystyle S_f^{(1)}=\sum_{l_1=0}^{l_1=J_1} \cos \frac{\pi n_1
(2 l_1+1)}{J_1+1} \cos \frac{\pi n_2}{J_2+1}\cos \frac{\pi n_3 (2
l_1+1)}{J+1} +(1\leftrightarrow 2),\end{array}\ee
\be \displaystyle S_g^{(1)}=\sum_{l_1=1}^{l_1=J_1} \cos \frac{\pi
n_1 (2 l_1+1)}{J_1+1} \cos \frac{\pi n_2}{J_2+1}\cos \frac{\pi
n_3(2 (l_1-1)+1)}{J+1} +(1\leftrightarrow 2), \hphantom{++}\ee
\be \displaystyle S_h^{(1)}=\cos \frac{\pi n_1}{J_1+1} \cos
\frac{\pi n_2}{J_2+1}\cos \frac{ \pi n_3}{J+1} +(1\leftrightarrow
2).\ee
All these contributions carry also the overall factor
$\frac{\lambda}{16\pi^2}$ from the one-loop interaction. They will
also carry numerical factors $c_i$ coming from the loop
integration. The loop integration and the resulting divergency
structure entering diagrams $(a)$ and $(e)$ is fully identical to
the one done for the two-point functions
in\mycite{Minahan:2002ve}; the diagrams $(d)$ and $(e)$ yield
twice as less divergency as $(a)$ or $(c)$. These factors are then
found from Table\myref{tab1}
$c_a=1,c_b=-1,c_c=1,c_d=1/2,c_e=1/2,c_f=1,c_g=1,c_h=1$. The total
one loop contribution will be then given by \beq
S^{(1)}=\frac{\lambda}{16\pi^2}\sum S^{(1)}_i c_i.\eeq
Summing everything up we get
\beq S^{(1)}=-\frac{4 n^3 \left( n_1^2 (1-y)^2 +n_2^2 y^2 +n_3^2
y^2 (1-y)^2 \right)\sin^2(\pi n_3
y)}{\left(n_3^2y^2-n_1^2\right)\left(n_3^2(1-y)^2-n_2^2\right)}
\eeq
whence we get
\beq
c^1_{123}=\frac{1}{\lambda^\prime}\frac{S^{(1)}}{S^{(0)}}=-\frac{1}{4}\left(\frac{n_1^2}{y^2}
+\frac{n_2^2}{(1-y)^2}+n_3^2 \right).\eeq
exactly as in the string theory above.
\section{Discussion}
We have observed that a three-point correlation function for all
dynamical BMN operators matches precisely the perturbative
weakly coupled planar field theory and the Penrose limit of the
strongly coupled string field theory at one loop level in the
Frolov--Tseytlin limit. This result is quite unexpected, since,
on one hand, a correlator of two heavy and one light operators
has been previously demonstrated in\mycite{Bissi:2011ha} to
 fail to match the semiclassical string calculation
in the Frolov--Tseytlin limit. On the other hand, a
heavy-heavy-light correlator calculated via integrability has been
shown to beautifully agree with the string theory in the
Frolov--Tseytlin limit, yet only in the thermodynamical regime,
when the number of excitations tends to infinity~\cite{Gromov:2012vu}. Our result is
thus the only one-loop analytic calculation of a three-point function so
far, where complete agreement between fields and strings is
observed. It has been noted in\mycite{Bissi:2011ha} that such a
matching is not necessarily present even at one-loop level, since
$C_{123}$ is unprotected. Thus our case should be
considered as another ``wonder'' of AdS/CFT and must be explained
somehow. The well known state/operator identification for BMN
states, which in other cases is not so well
established~\cite{Bissi:2011dc,Bissi:2011ha}, certainly helps in
providing this matching. However, we do not yet possess a generic
argument why this must work in a more general setting;
neither we can guess which corner of the parameter space may be
covered by the conjecture on exact matching between the
three-point functions on gauge and gravity sides. The Penrose
limit string field theory Hamiltonian which is the basis of our
string calculation seems to know nothing about the Yang-Mills
planar correspondence, yet it reproduces its results
astonishingly. On the other hand, Yang-Mills planar theory does
indeed produce the terms of order $\lambda J^2$, which were
supposed by authors of\mycite{Bissi:2011ha} to be exactly the
stumbling block in the heavy-heavy-light matching. In the
three-BMN case these stumbling blocks cancel each other
accurately, leaving only terms of similar orders both in $\lambda$
and in $J$ on both sides of the correspondence.

\section*{Acknowledgments}

We thank Agnese Bissi, Marta Orselli and Kolya Gromov  for
interesting and stimulating discussions.  This work was supported
in part by the MIUR-PRIN contract 2009-KHZKRX. The work of A.Z. is
supported in part by the Ministry of Education and Science of the
Russian Federation under contract 14.740.11.0081, NSh 3349.2012.2,
the RFBR grants 10-01-00836 and 10-02-01483.


\providecommand{\href}[2]{#2}\begingroup\raggedright\endgroup

\end{document}